\documentclass[10pt,conference]{IEEEtran}

\usepackage{makecell}
\usepackage{comment}
\usepackage{tcolorbox}
\usepackage{caption}
\usepackage{color,xcolor}
\usepackage{xspace}
\usepackage{multirow}
\usepackage{booktabs}
\usepackage{lipsum}
\usepackage{soulutf8}
\usepackage{diagbox}
\usepackage{physics}

\usepackage{graphicx}
\usepackage{colortbl}

\usepackage{enumitem}
\usepackage{braket}
\usepackage{calc}

\usepackage{amsmath}
\usepackage{amsfonts}
\usepackage{url}

\usepackage{afterpage}
\usepackage{cancel}

\usepackage[ruled,vlined]{algorithm2e}
\usepackage{setspace}

\usepackage{pifont}

\usepackage{titlesec}

\usepackage{hyperref}
\hypersetup{
    colorlinks=true,
    linkcolor=magenta,
    filecolor=magenta,      
    urlcolor=blue,
}

\usepackage{xcolor}
\usepackage{soul}

\newcommand{\name}{ConiQ}

\newcommand{\mhcc}{many-hypercube codes}
\newcommand{\Mhcc}{Many-hypercube codes}
\newcommand{\MHCC}{Many-hypercube Codes}

\newcommand{\aha}{AHA}
\newcommand{\ahafull}{Automorphism-assisted Hierarchical Addressing}
\newcommand{\vair}{VAIR\xspace}
\newcommand{\vairfull}{Virtual Atom Intermediate Representation}

\newcommand{\defn}[1]{\emph{#1}}
\newcommand{\movp}[2]{$\mathbf{Mov}(#1, #2)$}
\newcommand{\mov}[3]{$\mathbf{Mov}^{#1}(#2, #3)$}
\newcommand{\transas}[3]{$\mathbf{Trans}_{\text{a}\to \text{s}}^{#1}(#2, #3)$}
\newcommand{\transsa}[3]{$\mathbf{Trans}_{\text{s}\to \text{a}}^{#1}(#2, #3)$}
\newcommand{\transasp}[2]{$\mathbf{Trans}_{\text{a}\to \text{s}}(#1, #2)$}
\newcommand{\transsap}[2]{$\mathbf{Trans}_{\text{s}\to \text{a}}(#1, #2)$}
\newcommand{\onergadget}[4]{$\mathbf{1Rgadget}^{#1}(#2,#3,#4)$}
\newcommand{\tworgadget}[4]{$\mathbf{2Rgadget}^{#1}(#2,#3,#4)$}
\newcommand{\oneqgate}[3]{$\mathbf{1Qgate}(#1,#2,#3)$}
\newcommand{\twoqcz}[2]{$\mathbf{ParallelCZ}(#1,#2)$}

\newcommand{\cz}{\texttt{CZ}\xspace}
\newcommand{\cx}{\texttt{CNOT}\xspace}
\newcommand{\swap}{\texttt{SWAP}\xspace}
\newcommand{\hgate}{\texttt{H}\xspace}

\newcommand{\xgate}{\texttt{X}\xspace}
\newcommand{\zgate}{\texttt{Z}\xspace}
\newcommand{\tgate}{\texttt{T}\xspace}

\definecolor{titleblue}{HTML}{003076}
\definecolor{backgroundblue}{HTML}{eef6fc}
\newtcolorbox{hintbox}[2][]
{
  colframe = titleblue!100,
  colback  = backgroundblue!100,
  boxsep=2pt,
  width=\dimexpr\columnwidth\relax, 
  coltitle = titleblue!20!black,
  title    = #2,
  #1,
}
\usepackage[capitalise]{cleveref}

\def\BibTeX{{\rm B\kern-.05em{\sc i\kern-.025em b}\kern-.08em
T\kern-.1667em\lower.7ex\hbox{E}\kern-.125emX}}
\begin{document}

\title{\name{}: Enabling Concatenated Quantum Error Correction on
Neutral Atom Arrays}

\makeatletter
\newcommand{\linebreakand}{
  \end{@IEEEauthorhalign}
  \hfill\mbox{}\par
  \mbox{}\hfill\begin{@IEEEauthorhalign}
}
\makeatother
\author{
  \IEEEauthorblockN{Pengyu Liu}
\IEEEauthorblockA{\textit{Computer Science Department} \\
\textit{Carnegie Mellon University}\\
Pittsburgh, PA \\
pengyuliu@cmu.edu}
\and
\IEEEauthorblockN{Mingkuan Xu}
\IEEEauthorblockA{\textit{Computer Science Department} \\
\textit{Carnegie Mellon University}\\
Pittsburgh, PA \\
mingkuan@cmu.edu}\and
\IEEEauthorblockN{Hengyun Zhou}
\IEEEauthorblockA{\textit{QuEra Computing Inc} \\
Boston, MA \\
hyharryzhou@gmail.com}\linebreakand
\IEEEauthorblockN{Hanrui Wang}
\IEEEauthorblockA{\textit{Computer Science Department} \\
\textit{University of California, Los Angeles}\\
Los Angeles, CA \\
wang@cs.ucla.edu}\and
\IEEEauthorblockN{Umut A. Acar}
\IEEEauthorblockA{\textit{Computer Science Department} \\
\textit{Carnegie Mellon University}\\
Pittsburgh, PA \\
umut@cmu.edu}\and
\IEEEauthorblockN{Yunong Shi}
\IEEEauthorblockA{\textit{Amazon Web Services} \\
New York, NY \\
shiyunon@amazon.com}
}

\maketitle
\begin{abstract}
  Recent progress on concatenated codes, especially \mhcc{}, achieves
  unprecedented space efficiency. Yet two critical challenges persist
  in practice.
  First, these codes lack efficient implementations of addressable
  logical gates. Second, the required high degree of parallelism and
  long-range interactions pose significant challenges for
  current hardware platforms.
  In this paper, we propose an efficient compilation approach for
  concatenated codes, specifically
  \mhcc{}, targeted at neutral atom arrays, which provide the
  necessary parallelism and long-range interactions.
  Our approach builds on two key innovations. First, we
  introduce \ahafull{} (\aha{}) logical $\cx{}$ gates that significantly reduce
  spacetime overhead compared to conventional distillation-based
  methods. Second,
  we develop \vairfull{} (\vair{}) that enables level-wise
  optimization and legalization. We
  implement these innovations in \name{}, a hardware-aware quantum compiler
  designed to compile fault-tolerant quantum circuits for neutral atom arrays
  using \mhcc{}. Our evaluation demonstrates that \name{} achieves
  up to $2000\times$ reduction in spacetime overhead and up to
  $10^6\times$ reduction in compilation time compared to
  state-of-the-art compilers, with our \aha{} gates providing an additional overhead reduction of up to $20\times$. These results establish
  concatenated codes as a promising approach for fault-tolerant
  quantum computing
  in the near future.
\end{abstract}

\section{Introduction}
Quantum computing inherently suffers from noise, making quantum error
correction (QEC) indispensable for practical quantum computation. Surface codes
have long been the leading candidate due to their high error threshold, but
their low encoding rate imposes substantial qubit overhead. For instance,
achieving a $10^{-9}$ logical error rate at a physical error rate of $10^{-3}$
requires a surface code of distance 15, incurring $225\times$ space
overhead~\cite{gidney2021factor}. Quantum Low-Density Parity Check
(qLDPC)
codes~\cite{panteleev2022asymptotically,dinur2023good,hastings2021fiber},
despite promising asymptotic properties, face significant near-term
limitations, including poor hardware adaptability and insufficient
understanding of practical logical gate
implementations~\cite{gottesman2022opportunities}.
\begin{figure}[t]
  \centering
  \includegraphics[width=1.\columnwidth]{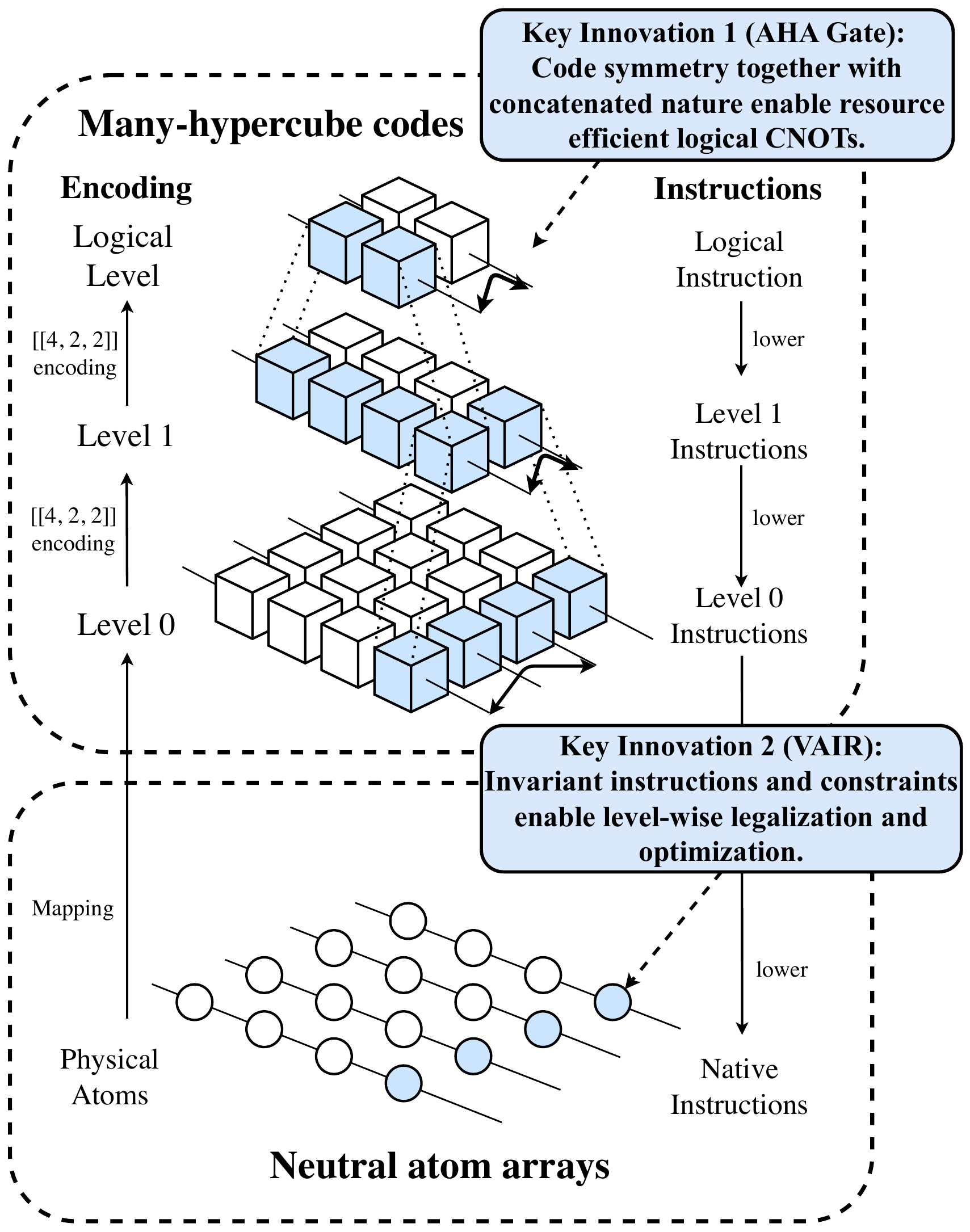}
  \caption{Overview of \name{} and its key innovations: (1) The
    \ahafull{} (\aha{}) scheme for efficient individually addressable
    logical gates. (2) The \vairfull{} (\vair{}) for efficient
  optimization and legalization across concatenation levels.}
  \label{fig:intro}
  \vspace*{-0.5cm}
\end{figure}

Recently, concatenated QEC codes have emerged as compelling alternatives due to
their impressive encoding rates. For example, \mhcc{}~\cite{goto2024many} represent a family of QEC codes constructed by
recursively concatenating quantum error-detecting codes with minimal sizes and
high qubit efficiency, such as the [[6,4,2]] codes (encoding 4 logical qubits
into 6 physical qubits with distance 2). This approach enables a substantial
reduction in physical qubit overhead. Four levels of concatenation yield an
encoding rate of $(4/6)^4 \approx 20\%$, allowing \mhcc{} to
encode 1 logical qubit using approximately 5 physical qubits on
average at a code
distance of 16, significantly outperforming both surface codes and qLDPC codes.
However, two critical obstacles hinder the practical deployment of concatenated
codes. First, current proposals rely on distillation for
individually addressable logical gates, particularly $\cx$ gates, which incur
considerable spacetime overhead~\cite{goto2024many}.
Second, the requirements for long-range interaction and high parallelism pose
substantial challenges for current hardware and compilers. These limitations
have thus far restricted concatenated codes to theoretical study, impeding
their application in real-world quantum systems.

The second challenge can potentially be addressed by leveraging neutral atom
arrays as the hardware platform. Neutral atom arrays offer numerous advantages
for fault-tolerant quantum computing: recent experiments demonstrate their long
coherence times, impressive scalability, and low-error
operations~\cite{manetsch2024tweezer,bluvstein2022quantum,Graham2022,evered2023high}.
Most importantly, the unique capability of neutral atom arrays to dynamically
reposition qubits during computation provides long-range interactions and
extensive parallelism—properties particularly well-suited for concatenated
codes. However, the long-range interactions and parallelism are in a
restricted form, and require special compilation techniques to
exploit the capabilities provided by neutral atom arrays. Although
compilers for neutral atom arrays such as
Atomique~\cite{wang2024atomique} and Enola~\cite{tan2025compilation}
exist, they focus on the compilation of general quantum circuits,
and result in prohibitive overheads in the context of
concatenated codes.

To harness the advantages of concatenated QEC codes and neutral
atom arrays, we introduce \name{}, a novel compiler framework specifically
designed for efficiently implementing concatenated QEC codes on neutral atom
arrays. \name{} builds upon two key innovations with an overview shown in
\cref{fig:intro}:

(1) By leveraging code symmetry and concatenation, we develop an \ahafull{}
(\aha{}) logical $\cx$ gate scheme that is individually addressable, requiring
only a few error correction cycles of overhead, dramatically improving
efficiency compared to prior distillation-based schemes.

(2) We introduce a \vairfull{} (\vair{}) that virtualizes logical registers as
physical atoms across concatenation levels. Crucially, we prove that \vair{}
preserves consistent program states, instruction sets, and hardware constraints
at each concatenation level, enabling efficient level-wise
legalization and optimization.

Compared to the state-of-the-art compilers
Atomique~\cite{wang2024atomique} and Enola~\cite{tan2025compilation} with
prior distillation-based $\cx$ gates, \name{} achieves over $10^4\times$
improvement in spacetime overhead and $10^6\times$ reduction in compilation
time. Thus, \name{} realizes the potential of high-encoding-rate quantum error
correction with dramatically reduced overhead, enabling practical deployment of
concatenated QEC codes on near-term neutral atom quantum computers.

\section{Background}

\subsection{Neutral Atom Arrays}
\label{sec:naa}

Neutral atom arrays possess the unique capability to move qubits during
computation, providing effective long-range connectivity and massive
parallelism. Here we formally describe the neutral atom array architecture,
program states, instruction set, and constraints at the physical level.
In~\cref{sec:parallel_gadgets}, we extend these concepts to the logical level
and design the \vairfull{} (\vair{}) for efficiently compiling concatenated
codes.

\noindent\textbf{Architecture.} Neutral atom arrays
utilize individually-focused laser spots to trap and manipulate
atoms. Two types of traps are typically employed: fixed spatial light modulators
(SLM) and movable acousto-optic deflectors (AOD). Different lasers
are typically used for SLM and AOD traps, so different type of traps
won't interfere with each
other. Each type of trap forms a two-dimensional square grid.
Atoms can reside in
either type of laser trap. AOD atoms can be dynamically repositioned by
adjusting the corresponding lasers. Conceptually, this configuration
creates two planar atom layers: a stationary SLM layer and a movable
AOD layer, slightly offset along the $z$ axis.

\begin{figure}[t]
    \centering
    \includegraphics[width=\columnwidth]{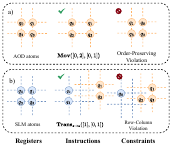}
    \caption{Neutral atom array architecture, instructions, and constraints: (a) An allowed move operation (left) and a forbidden move operation that violates the order-preserving constraint (right). (b) An allowed transfer operation affecting entire rows and columns (left) and a forbidden operation that violates the row-column constraint (right).}
    \label{fig:constraints}
    \vspace*{-0.5cm}
\end{figure}

\noindent\textbf{Program State Representation.} We
represent the collective system state, particularly atom positions, as
a four-tuple $(I,J,A,S)$:

\begin{itemize}
  \item $I$: Ordered x-coordinates of AOD traps. $I[i]$ denotes the
    $x$ coordinate of the $i$-th column, where $I[i]<I[i']$ for $i<i'$.
  \item $J$: Ordered y-coordinates of AOD traps. $J[j]$ denotes the
    $y$ coordinate of the $j$-th row, where $J[j]<J[j']$ for $j<j'$.
  \item $A$: Positions of AOD atoms, where $A[i][j]$ identifies the
    atom at the $i$-th column and $j$-th row, located at
    $(I[i],J[j])$ or is empty $(\emptyset)$ when no atom is present.
  \item $S$: Positions of SLM atoms on an integer grid, where
    $S[i][j]$ identifies the atom index at integer coordinates
    $(i,j)$ (assuming SLM atoms are placed in a unit grid) or is
    empty $(\emptyset)$ when no atom is present.
\end{itemize}

\noindent\textbf{Instruction Set.} Neutral atom arrays
support quantum gate operations as well as parallel atom transfer and
movement through the following instruction set. Here, we use $I'$ and $J'$ to represent subsets of $I$ and $J$, respectively:
\begin{itemize}
  \item \transasp{I'}{J'} and \transsap{I'}{J'}: Transfer atoms
    between SLM and AOD traps at intersections of rows $I'$ and
    columns $J'$. $a\to s$ indicates transfer from AOD to SLM and
    vice versa. For \transsap{I'}{J'}, if an SLM atom exists at
    $S[i'][j']=n$ where $i'\in I', j'\in J'$, and there is an empty AOD trap
    $A[i][j]=\emptyset$ where $I[i]=i'$ and $J[j]=j'$ (the SLM atom
    and the empty AOD trap overlap in the $x$-$y$ plane), then the atom is
    transferred by setting $A[i][j]=n$ and $S[i'][j']=\emptyset$.
    \transasp{I'}{J'} performs the reverse transfer.
  \item \movp{I}{J}: Relocates AOD atoms to new positions, updating
    the state to $(I,J,A,S)$. The movement must satisfy the
    order-preserving constraint: $I$ and $J$ remain ordered.
  \item \oneqgate{U}{I'}{J'}: Applies single-qubit operations $U$ to SLM atoms at
    specified grid coordinates. For all $i'\in I', j'\in J'$, if there is
    an SLM atom $S[i'][j']=n$, then $U$ is applied to qubit $n$.
  \item \twoqcz{I'}{J'}: For all SLM atoms $S[i'][j']$, where $i'\in I',
    j'\in J'$, if there is an AOD atom overlapping with it, a $\cz$
    gate is applied between the SLM atom and the AOD
    atom.\footnote{In current neutral atom arrays, $\cz$ gates are
      implemented globally. The \twoqcz{I'}{J'} operation can be realized using a
      global $\cz$ gate with two move operations~\cite{wang2024atomique}.
      While one-qubit gates can also be applied to AOD atoms, and
      two-qubit gates between SLM-SLM or AOD-AOD pairs are possible,
      we omit these
    capabilities in this paper for simplicity.}
\end{itemize}

\noindent\textbf{Program Constraints.} The
instruction set for neutral atom arrays offers significant parallelism but
with specific constraints (illustrated in \cref{fig:constraints}):
\begin{itemize}
  \item \textit{Order-Preserving Constraint}: Move instructions must
    maintain the relative order of AOD atoms to prevent collisions
    (\cref{fig:constraints}~(a)).
  \item \textit{Row-Column Constraint}: Instructions are executed on
    all intersections of entire rows and columns (\cref{fig:constraints}~(b)).
\end{itemize}
While available instructions provide parallelism that could potentially be
advantageous for concatenated codes, the constraints also pose
significant challenges for efficient circuit compilation: unlike in
superconducting qubits, where logically independent physical
operations acting on neighboring qubits can always be scheduled in parallel.

\subsection{QEC Basics and Many-Hypercube Codes}
\label{sec:qec_basics}

We will focus on \mhcc{} in this paper, which are the leading candidate in
concatenated QEC codes, but our approach, especially the \vair{}
compilation, can be applied to other concatenated QEC codes.

\noindent\textbf{Error Detecting Codes.} Error detecting codes are QEC
codes capable of detecting errors but not correcting them. The simplest
family of error-detecting codes is the $[[2m,2m-2,2]]$
code~\cite{ErrorCorrectionZoo}, which encodes $2m-2$ logical qubits into
$2m$ physical qubits with code distance $2$. We denote this family as
$D_{2m}$. The $D_{2m}$ code has two stabilizers:
$\xgate_1\cdots\xgate_{2m}$ and $\zgate_1\cdots\zgate_{2m}$. As an
example, \cref{tab:d4_logical} presents the logical operators of $D_4$,
the smallest instance of the $D_{2m}$ family.

\noindent\textbf{Automorphism of codes.} Automorphisms of codes are
symmetries that can be exploited to construct efficient fault-tolerant
gadgets through physical qubit permutation. Take the $D_4$ code as an
example: by swapping the
first and third physical qubits, we exchange the logical operators of
the two logical qubits~\cite{knill2007quantum}, which creates a
logical $\swap$ gate. This automorphism
property will be leveraged in
our $\cx{}$ gate implementation.

\begin{table}[h]
\centering
\begin{tabular}{c|cccc}
\toprule
Logical Op. & $q_1$ & $q_2$ & $q_3$ & $q_4$ \\
\midrule
$\xgate_{L_1}$   & $\xgate$   & $\xgate$   &       &       \\
$\xgate_{L_2}$   &       & $\xgate$   & $\xgate$   &       \\
\midrule
$\zgate_{L_1}$   &       & $\zgate$   & $\zgate$   &       \\
$\zgate_{L_2}$   &       $\zgate$   & $\zgate$   &&       \\
\bottomrule
\end{tabular}
    \caption{Logical operators of the $D_4$ error detecting code. Each row shows the combination of physical operators that implement a logical operation. For example, applying $\xgate$ gates to physical qubits 1 and 2 implements an $\xgate$ operation on the first logical qubit.}

    \label{tab:d4_logical}
    \vspace*{-0.3cm}
\end{table}

\noindent\textbf{\MHCC{}.} The $D_{2m}$ code alone cannot correct
errors and is therefore unsuitable for fault-tolerant quantum
computation. Concatenation offers a powerful technique for constructing
codes with enhanced error correction capabilities from simpler codes.
The concatenated $D_{6}$ codes are known as \mhcc{}~\cite{goto2024many}.
Concatenation encodes physical qubits using an
initial code, then treats the resulting logical qubits as physical inputs for
the next encoding level. This process can be repeated to enhance error
correction capabilities. We use \defn{level} to denote each concatenation
iteration, and \defn{register} to refer to a block of logical qubits treated
collectively at a particular level.

In this paper, we extend the definition of \mhcc{} to include both
$D_6$ and $D_4$ codes and denote an $l$-level many-hypercube code as
$D_{n_1,n_2,\dots,n_l}$, indicating
that level-$i$ uses a $D_{n_i}$ code. For example, the code $D_{4,4,6,6}$
employs two levels of $D_4$ (near the physical level), followed by two levels
of $D_6$ encoding (near the logical level). We use $q_{i_1\cdots
i_l}$ to denote the $i_l$-th logical qubit in the code consisting of
the level-$(l-1)$ logical qubits $q_{i_1\cdots i_{l-1}}$.

\cref{fig:concat_example} illustrates a $D_{4,4}$ code with 16 physical
qubits. First, groups of four qubits are encoded using the $D_4$ code, creating
level-1 registers with 2 logical qubits each. Then the two columns of level-1
logical qubits are encoded using the $D_4$ code separately, resulting
in a level-2
register with four logical qubits. This demonstrates why we refer to the four
logical qubits as a register: the information of these logical qubits
is interleaved across the physical qubits and must be processed collectively.
In this example, logical Pauli gates can be applied based on
\cref{tab:d4_logical}. As shown in the figure, to apply a logical
$\xgate$ gate to the first logical qubit in a level-$2$ register, we apply two
level-$1$ $\xgate$ gates, which are further decomposed into four
physical $\xgate$ gates.

\begin{figure}[t]
	\centering
	\includegraphics[width=\columnwidth]{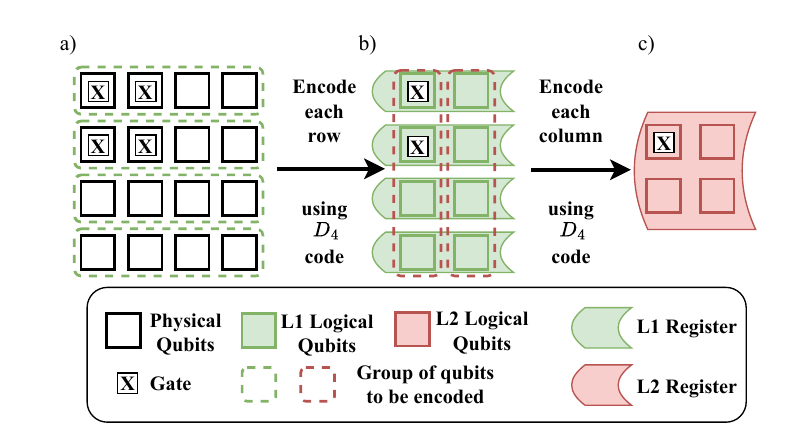}
	\caption{Construction of a level-2 $D_{4,4}$ many-hypercube code: (a) Initial arrangement of physical qubits. (b) First level of concatenation. (c) Second level of concatenation.}
	\label{fig:concat_example}
	\vspace*{-0.3cm}
\end{figure}

\begin{figure}[ht]
    \centering
    \includegraphics[width=0.7\columnwidth]{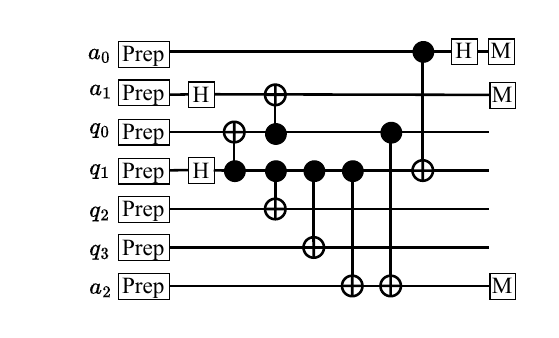}
    \caption{State preparation gadget for a level-2 register in $D_{4,4}$ code.}
    \label{fig:encoding}
    \vspace*{-0.5cm}
\end{figure}

\noindent\textbf{Fault-tolerant Gadgets of \MHCC{}.} Fault-tolerant
gadgets represent atomic instructions at the logical level that
guarantee the desired fault-tolerant
properties~\cite{gottesman2002introduction}.
We focus on
two types of gadgets: state preparation and logical instructions.

\noindent\textit{State Preparation.} State preparation constitutes the
first step of a fault-tolerant circuit and is essential for other
logical instructions. \cite{goto2024many} proposed a state preparation
gadget that offers an effective trade-off between logical error rate
and overhead. Taking the level-2 state preparation protocol as an example,
as illustrated in \cref{fig:encoding}, the inputs $q_0$ to $q_3$ and $a_0$ to
$a_2$ are all level-1 registers. These registers first undergo a
level-1 encoding
circuit, followed by a series of transversal gates. Conditioned on the
measurement results of the ancilla registers, we can prepare the
level-2 encoded state fault-tolerantly. From this example, we observe two
critical properties of \mhcc{} gadgets: (1) Higher-level gadgets often require
identical copies of lower-level gadgets that can, in principle, be
batched—a property that should be exploited during compilation. (2)
Numerous long-range interactions are required, which may be expensive
to implement.

\noindent\textit{Logical Instructions.} Since each code block in the
$D_{2m}$ family encodes multiple logical qubits in one register, the
ability to individually address specific logical qubits within a block
is crucial for computation. The conventional approach for implementing
addressable logical $\cx{}$ gates relies on plus state
distillation~\cite{goto2024many}, which requires exponentially increasing
resources. However, the $D_{2m}$ code, being a CSS code, permits simpler
implementations of certain gates. Logical $\cx{}$ gates between all
corresponding
logical qubits of two registers can be implemented transversally by applying
physical $\cx{}$ gates between all corresponding physical-qubit
pairs~\cite{calderbank1996good}. Additionally, the $D_{2m}$ codes feature
transversal $\hgate$ gates (up to a permutation of the logical qubits).

\section{Challenges in Practical Implementations of Concatenated Codes on Neutral Atom Arrays} \label{sec:challenge}

Although concatenated QEC codes offer excellent space efficiency as
quantum memory, two key challenges hinder their practical deployment:
(1) Existing logical gates, particularly individually addressable gates, incur
significant spacetime overhead for computation. (2) The absence of efficient
compilation methods prevents effective utilization of the intrinsic
parallelism in neutral atom arrays.
We will examine these challenges in detail in the following sections.

\subsection{Efficient Addressable Logical Gate Implementation}

The current distillation-based approach for implementing individually
addressable logical gates introduces
significant overhead, primarily because state distillation must be
repeated at each level of
concatenation~\cite{goto2024many,yamasaki2024time}. At every level,
high-fidelity resource states are generated by consuming multiple
noisy copies from the lower level. This recursive structure compounds
the cost exponentially, causing overall resource
requirements to grow rapidly with concatenation levels. Consequently, it is
crucial to develop low-cost logical gate schemes for practical near-term
implementation.

\subsection{Effectively Leveraging Neutral Atom Arrays' Intrinsic Parallelism}

Compiling concatenated QEC codes typically involves two tasks: register
mapping and gadget scheduling. One approach is to fully unfold all
registers and gadgets to the physical level before performing mapping
and scheduling. Current compilers, such as
Atomique~\cite{wang2024atomique} and Enola~\cite{tan2025compilation}, follow this method. However, this
approach neglects the hierarchical structure inherent in concatenated
QEC codes, resulting in significant spacetime and compilation overhead.

A more effective compilation strategy maps registers and schedules
gadgets level-by-level, closely aligning with the hierarchical
structure of concatenated QEC codes. However, current compilation methods
lack a program representation that efficiently exposes physical
constraints of neutral atom arrays to higher logical levels, making
optimization (batching parallelizable gadgets) and
legalization (ensuring physical-level feasibility) nearly impossible.

Without properly exposing these physical constraints, only two naive
strategies remain: (1) prioritizing legalization by sequentially
scheduling all gadgets, thus severely underutilizing hardware
parallelism, or (2) prioritizing optimization by batching all
logically independent gadgets, potentially generating illegal
schedules that must be resolved at the physical level.

Both strategies lead to catastrophic consequences—ignoring
legalization produces unsound and unusable compilation outcomes,
whereas sacrificing parallelism causes exponential runtime growth due to multiplicative inefficiencies compounding across concatenation levels.
We call this exponential inefficiency \textit{cascading latency amplification}.

Therefore, it is crucial to develop a compilation method capable of
level-by-level compilation that simultaneously achieves legalization
and optimization by clearly exposing allowed instructions and
physical constraints at every concatenation level.

\section{\name{} Architecture}
\label{sec:architecture}
\begin{figure}[t]
  \centering
  \includegraphics[width=\columnwidth]{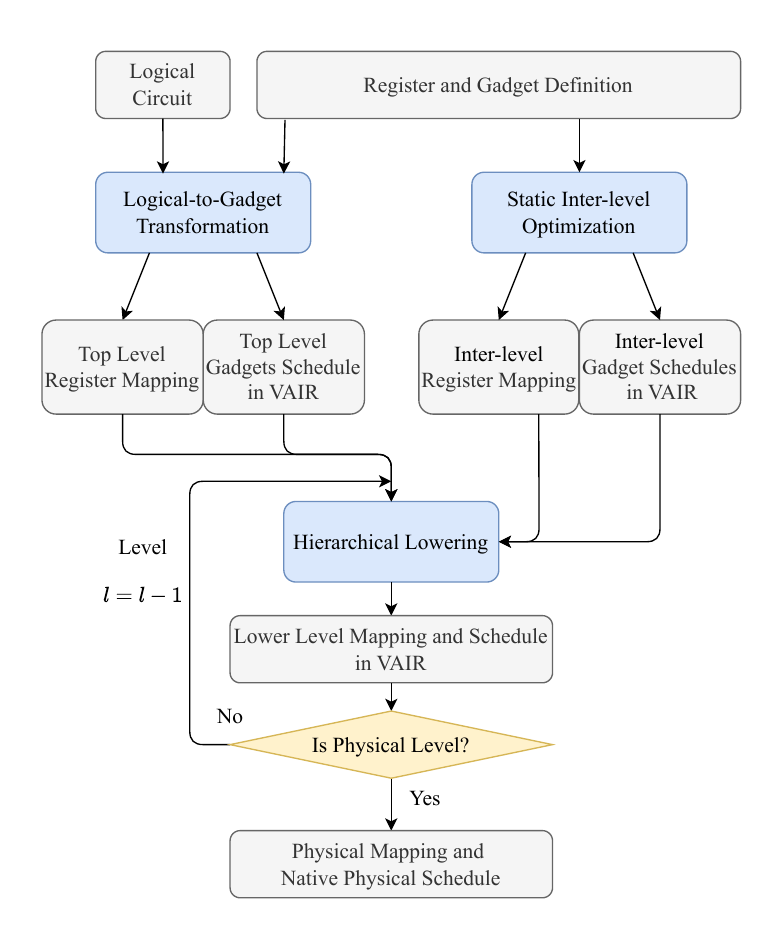}
  \caption{Three-phase compilation workflow of \name{}.
  }
  \label{fig:compilation_overview}
  \vspace*{-0.5cm}
\end{figure}

\name{} is specifically designed to address the challenges outlined
above through two key innovations:
\begin{enumerate}
  \item \textbf{\ahafull{} (\aha{}) Gates}: By leveraging the symmetry and
    concatenated structure of codes, \name{} introduces an \ahafull{}
    logical gate scheme, significantly reducing overhead compared to
    distillation-based methods.

  \item \textbf{\vairfull{} (\vair{})}: At the core of \name{} is
    \vair{}, a hierarchical intermediate representation explicitly
    designed to expose allowed instructions and physical constraints
    of neutral atom arrays to each concatenation level. We prove that
    scheduling instructions compliant with \vair{} constraints at
    higher levels directly ensures both legality and
    parallelism at the physical level.
\end{enumerate}
Built upon these key innovations, \name{} employs a structured, level-wise
compilation workflow, illustrated in~\cref{fig:compilation_overview}.
The workflow comprises three phases:

\noindent\textbf{(1) Logical-to-Gadget Transformation}:
This phase converts the input logical circuit into a sequence of
top-level fault-tolerant gadgets. We leverage our \aha{} to implement
addressable logical gates with minimized gadget cost while ensuring
fault tolerance requirements~\cite{gottesman2002introduction}.

\noindent\textbf{(2) Static Inter-level Optimization}:
Starting with gadget definitions for each concatenation level, this
phase generates templates of optimized register mappings and
schedules using the \vair{} framework. \vair{} enables independent
optimization and legalization of each level, ensuring that physical
runtime decreases
proportionally with each level's runtime reduction.

\noindent\textbf{(3) Hierarchical Lowering}:
Using the optimized templates produced in the previous phase,
hierarchical lowering recursively translates instructions from higher
levels down to physical instructions. \vair{} guarantees
that no additional overhead is introduced when compiling to lower levels and
that all physical instructions are native to the neutral atom arrays.

This structured, three-phase compilation not only exploits the
hierarchical structure of concatenated codes, but also reuses the
templates produced in the static optimization phase, significantly
reducing compilation overhead compared to existing compilers,
effectively unlocking the full potential
of concatenated QEC codes on neutral atom array hardware.

\section{\ahafull{} Gate Scheme}
\label{sec:efficient_logical_gates}
\begin{figure}[t]
    \centering
    \includegraphics[width=\columnwidth]{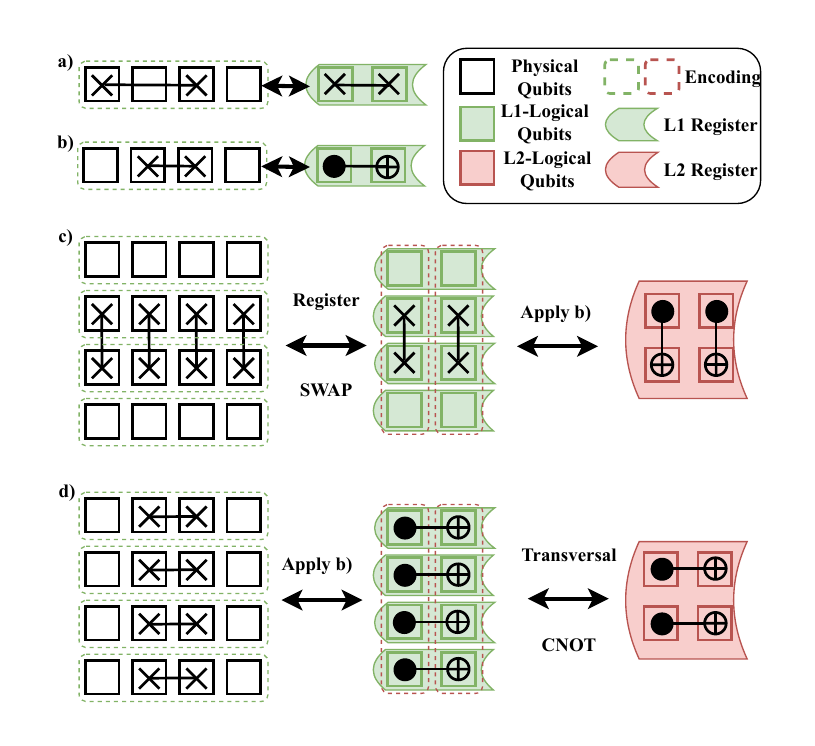}
    \caption{Implementation of logical operations through physical qubit permutations: (a) A logical $\swap$ gate implemented by swapping the first and third physical qubits. (b) A logical $\cx{}$ gate implemented by swapping the second and third physical qubits. (c) Intra-register batched $\cx{}$ gates achieved by swapping the second and third rows of physical qubits in a $4\times 4$ grid. (d) Inter-register batched $\cx{}$ gates along another dimension, implemented by swapping the second and third columns of physical qubits.}
    \label{fig:logical_op1}
    \vspace*{-0.6cm}
\end{figure}

In this section, we introduce the \ahafull{} (\aha{}) scheme, which
enables individually addressable logical instructions with
significantly reduced overhead. For clarity, we illustrate the
scheme using the $D_4$ code, though the methods readily extend to the $D_6$ code.

We present this section in two parts: first summarizing the
fundamental instructions required to implement the \aha{}
scheme, then detailing the construction process of individually addressable gates.

\noindent\textbf{Fundamental Instructions for \aha{}.}
The \aha{} scheme requires the following essential
logical instructions:

\begin{enumerate}
  \item State preparation and error correction.
  \item Transversal $\cx{}$ gates.
  \item Intra-register batched $\swap$ operations along a dimension: $\prod
    \swap(q_{i_1,\dots,0,\dots,i_l},q_{i_1,\dots,1,\dots,i_l})$.
  \item Intra-register batched $\cx{}$ operations along a dimension: $\prod
    \cx{}(q_{i_1,\dots,0,\dots,i_l},q_{i_1,\dots,1,\dots,i_l})$.
\end{enumerate}

The original \mhcc{} construction~\cite{goto2024many} directly
provides instructions 1--3. Instruction 4, critical for the \aha{}
scheme, is newly introduced in this work and detailed below.

\noindent\textbf{Intra-register Instructions via Automorphisms.}
As illustrated in \cref{fig:logical_op1}~(a, b)
and~\cref{sec:qec_basics}, the automorphisms of the $D_4$ code
enable intra-register logical $\swap$ and logical $\cx{}$ through
strategically selected physical $\swap$ operations~\cite{knill2007quantum}.
At higher concatenation levels, we generalize these patterns by
combining lower-level intra-register $\swap$ and $\cx{}$ operations, as depicted
in \cref{fig:logical_op1}~(c, d). By carefully selecting physical
qubit pairs for $\swap$, we can implement \defn{intra-register batched} logical $\cx{}$ gates
along any desired dimension, naturally scaling to
arbitrary concatenation levels.

\begin{figure}[t]
    \centering
    \includegraphics[width=\columnwidth]{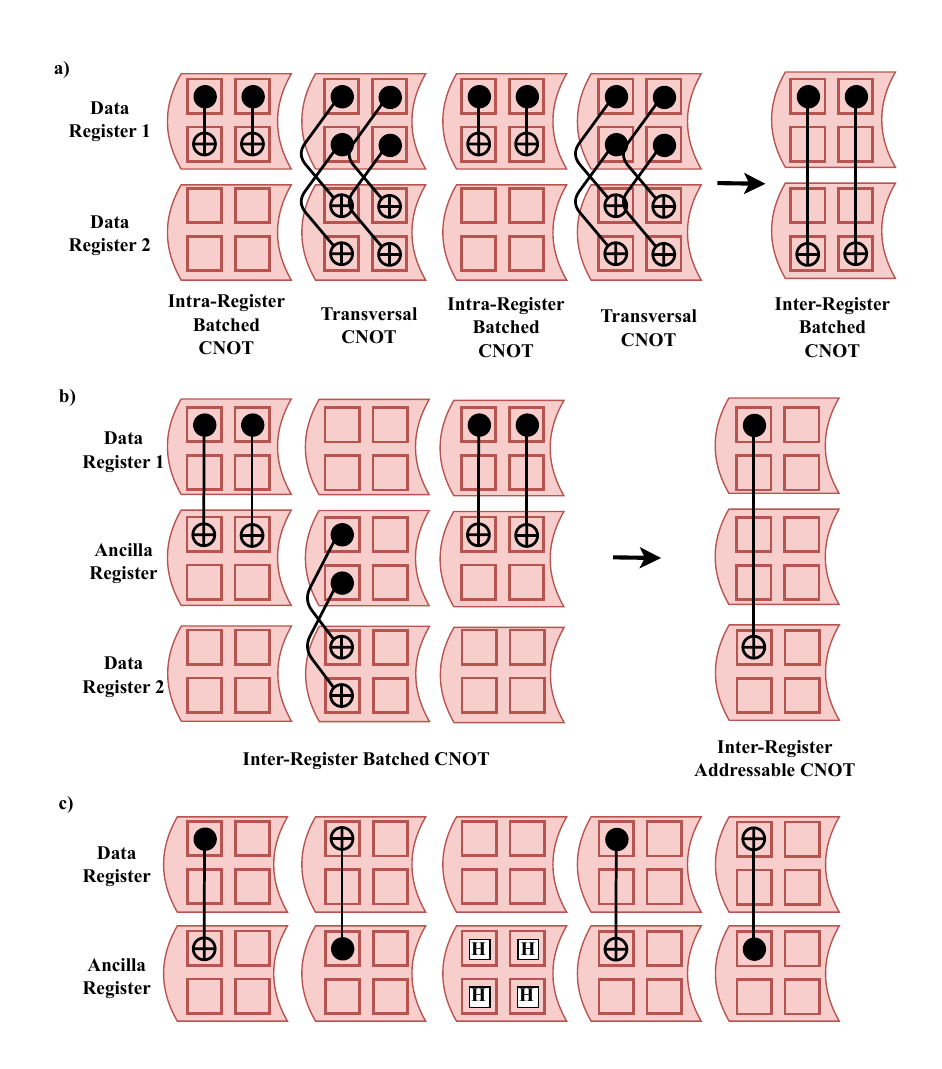}
    \caption{Implementation of individually addressable logical gates: (a) Inter-register batched $\cx{}$ gates on multiple logical qubits. 
    (b) Individually addressable $\cx{}$ gates achieved through hierarchical addressing. (c) Individually addressable $\hgate$ gates implemented through a combination of logical swaps and transversal operations. Note that error correction cycles are required for fault tolerance but are omitted here for clarity.}
    \label{fig:logical_op2}
    \vspace*{-0.5cm}
\end{figure}

\noindent\textbf{Addressable Logical $\cx{}$ via Hierarchical
Addressing.} With the fundamental intra-register operations
established, we next describe the construction of individually addressable logical $\cx{}$ gates using hierarchical addressing.
Although intra-register batched gates alone cannot address
individual logical qubits, they effectively separate logical qubits
into two distinct halves. To achieve full addressability, we
iteratively refine this division in a binary-search-like process
termed \textit{hierarchical addressing}.

We first construct the \defn{inter-register batched} $\cx{}$ gate, illustrated in
\cref{fig:logical_op2}~(a). This design is inspired by the bridge gate~\cite{itoko2020optimization}, which allows us to
selectively operate on half of the logical qubits. Using ancilla
registers, we continue this
selection process until we can address a single logical qubit, as shown in
\cref{fig:logical_op2}~(b). For higher concatenation levels, this selection
process can be repeated at each level of the hierarchy. Notably, $\swap$
and transversal $\cx{}$ operations can be performed in constant time with minimal
overhead. Thus, the cost of hierarchical addressing is primarily determined by
error correction.

\noindent\textbf{Other Addressable Logical Instructions.} Using the
addressable logical $\cx{}$ together with the transversal $\hgate$
gadget, we can efficiently implement addressable logical $\hgate$
on specific logical qubits, as illustrated in
\cref{fig:logical_op2}~(c).
Beyond Clifford instructions, non-Clifford instructions such as the $\tgate$ gate can
be implemented through magic state preparation and teleportation using
the addressable $\cx{}$ described above.

\section{\vair{}: \vairfull{}}
\label{sec:parallel_gadgets}

At the core of \name{} is the \vairfull{} (\vair{}), which enables
the efficient, level-wise compilation while ensuring legality and optimized
compilation outcomes. By explicitly modeling logical states,
instructions, and register constraints at every concatenation level,
\vair{} guarantees that any optimization or legalization performed at
higher levels inherently respects physical-level constraints and
preserves hardware parallelism.

We first define the \vair{} abstraction formally, and then detail its
key properties that enable level-wise compilation.

\subsection{\vair{} Definition}
\begin{figure}[t]
  \centering
  \includegraphics[width=\columnwidth]{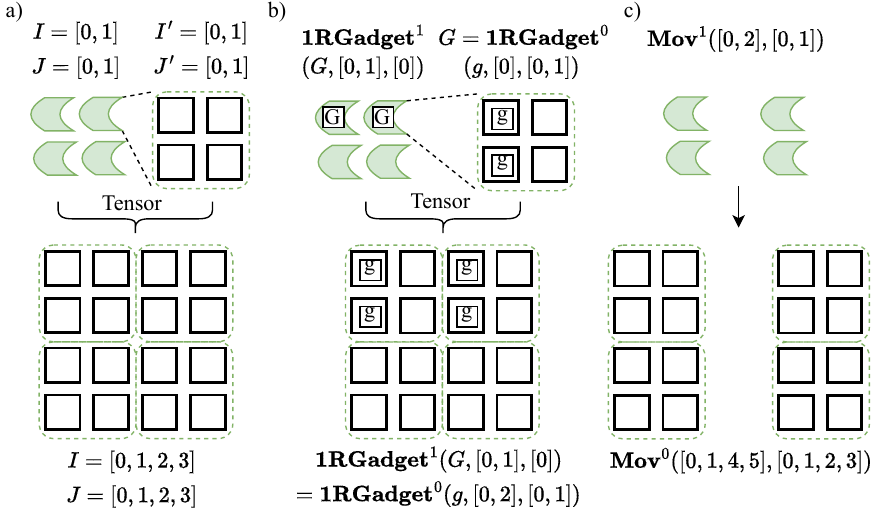}
  \caption{\vair{} lowering process. (a) Register mapping from four
    level-1 registers to their constituent level-0 registers. (b) An
    example showing how to implement level-1 gadgets using level-0
    gadgets. (c) An example showing how to implement level-1 movements
  using level-0 movements.}
  \label{fig:mapping}

  \vspace*{-0.5cm}
\end{figure}

\label{sec:nair}
\vair{} extends the neutral atom model to higher levels of abstraction, treating logical registers as virtual atoms to maintain consistent constraints across all levels.
At each level-$l$, the system state is represented by a four-tuple $\mathcal{R}^l=(I, J, A, S)$, defined as follows:

\begin{itemize}
  \item SLM registers are arranged on a unit-distance integer grid,
    with the $i$-th column and $j$-th row at $(i,j)$.
  \item $I$ and $J$ denote the ordered x- and y-coordinates of the
    AOD registers, respectively, maintaining the ordering constraints
    ($I[i]<I[i']$ for $i<i'$, and similarly for $J$).
  \item $A[i][j]$ maps the location of the AOD register to its index.
    $A[i][j]=n$ means the $n$-th register is of type AOD and lies at
    $(I[i], J[j])$. An empty position is denoted by $\emptyset$.
  \item $S[i][j]$ maps the location of the SLM register to its index.
    $S[i][j]=n$ means the $n$-th register is of type SLM and lies at $(i, j)$.
\end{itemize}

At every concatenation level-$l$, the supported instructions mirror
their physical counterparts at level-$0$. Here, we use $I'$ and $J'$ to represent subsets of $I$ and $J$, respectively:
\begin{itemize}
  \item \transas{l}{I'}{J'} and \transsa{l}{I'}{J'}: These operations transfer
    registers between AOD and SLM types. The subscript indicates the
    transfer direction ($a\to s$ for AOD to SLM, $s\to a$ for SLM
    to AOD). For \transsa{l}{I'}{J'}, if an SLM register exists
    where $S[i'][j']=n$ with $i'\in I', j'\in J'$, and there exists
    an empty AOD position $A[i][j]=\emptyset$ where $I[i]=i'$ and
    $J[j]=j'$, then the register is transferred by setting
    $A[i][j]=n$ and $S[i'][j']=\emptyset$. \transas{l}{I'}{J'}
    performs the inverse operation.
  \item \mov{l}{I}{J}: Moves AOD registers to new positions
    specified by $I$ and $J$, transforming the state to
    $(I,J,A,S)$. The operation must preserve ordering: $I[i] <
    I[i']$ for $i<i'$ and $J[j] < J[j']$ for $j<j'$.
  \item \onergadget{l}{G}{I'}{J'}: For $S[i'][j']=n$ where $i'\in I'$ and $j'\in J'$, apply the single-register gadget $G$ to register $n$.
  \item \tworgadget{l}{G}{I'}{J'}: Applies a two-register gadget $G$ between
    overlapping AOD-SLM register pairs specified by $I'$ and $J'$. Specifically, for any SLM register
    $S[i'][j']=m$ where $i'\in I'$ and $j'\in J'$, if there exists an
    AOD register $n$ that overlaps with $m$ in the $x$-$y$ plane,
    the operation $G(n,m)$ is applied.
\end{itemize}

\subsection{Level-wise Optimization and Legalization using \vair{}}
\label{sec:nair_realize}

The uniformity of the \vair{} abstraction across all concatenation
levels enables efficient and independent compilation at each level.
At level-$0$, \vair{} reduces to the physical neutral atom model,
guaranteeing physical feasibility of all compiled instructions. The
compilation process comprises two main steps at each level: register
mapping and gadget scheduling.

\noindent\textbf{Register Mapping.}
To map a single level-$l$ register to its constituent level-$(l-1)$
registers, we define a systematic coordinate-based approach. For a
single level-$l$ register consisting of $N$
subregisters with coordinates $(i'_k,j'_k)$, the $k$-th subregister
of a level-$l$ register at position $(i,j)$ is mapped to the coordinates:
\begin{align*}
  (i\cdot I_m^*+i_k',j\cdot
  J_m^*+j_k')
\end{align*}

We denote $I^*=\{i'_1,\dots,i'_N\}$, $J^*=\{j'_1,\dots,j'_N\}$, $I_m^*=\max(I^*)$ and $J_m^*=\max(J^*)$. This mapping is illustrated in
\cref{fig:mapping}~(a).

The entire mapping of level-$l$ registers can be concisely expressed
using tensor products:
\begin{align*}
I\otimes I^*=\{i\cdot I_m^*+i'|i\in I,i'\in I^*\}
\end{align*}

And similarly for $J$. This tensor product formulation allows us to describe a
coarse-grained representation of high-level registers without needing
to specify their lower-level implementations.

\name{} adopts an alternating linear mapping strategy, alternating between x
and y axes for different concatenation levels for its simplicity while producing an approximately
square physical layout.
\begin{algorithm}[h]
    \caption{Greedy Scheduling of Lower-Level Instructions}
    \label{alg:compile_gadgets}
    \KwIn{A sequence of level-$(l-1)$ instructions $[g_1, g_2, \ldots, g_n]$ that implements a higher-level gadget $G$}
    \KwOut{Optimized schedule $S$ of level-$(l-1)$ instruction represented in \vair{}}
    $S \leftarrow []$ \tcp*{Initialize empty schedule}
    $\text{remaining} \leftarrow [g_1, g_2, \ldots, g_n]$ \tcp*{Instructions to be scheduled}
    \While{$\text{remaining} \neq \emptyset$}{
        $\mathcal{F} \leftarrow \text{frontLayer}(\text{remaining})$ \tcp*{Extract instructions with no dependencies}
        $\text{instructionType} \leftarrow \text{randomSelect}(\mathcal{F})$ \tcp*{Randomly choose an instruction type}
        $\text{batchedSet} \leftarrow \text{maximalAddressableSet}(\mathcal{F}, \text{instructionType})$ \tcp*{Find largest set satisfying row-column constraint}
        $\text{thisInstruction} \leftarrow \text{generateInstruction}(\text{batchedSet}, \text{instructionType})$ \tcp*{Generate the instruction represented in \vair{}}
        $\text{append}(S, \text{thisInstruction})$ \tcp*{Add batched instruction to schedule}
        $\text{remaining} \leftarrow \text{remaining} \setminus \text{thisInstruction}$ \tcp*{Remove scheduled instructions}
    }
    \Return{$S$}
\end{algorithm}

\noindent\textbf{Gadget Scheduling.}
As a result of expressing register mappings as tensor products, each
level-$l$ instruction can be scheduled to level-$(l-1)$ instructions efficiently by the following rules:

\begin{itemize}
\item \transas{l}{I}{J} and \transsa{l}{I}{J}: These
  transfers are implemented as \transas{l-1}{I\otimes I^*}{J\otimes
  J^*}.

\item \mov{l}{I}{J}: Implemented as
  \mov{l-1}{I\otimes I^*}{J\otimes J^*}. We can readily verify
  that $I\otimes I^*$ and $J\otimes J^*$ satisfy the
  order-preserving constraint if the input $I$ and $J$ satisfy the
constraint. An example is illustrated in \cref{fig:mapping}~(c).

\item \onergadget{l}{G}{I}{J}: If $G$ is composed of $n$
level-$(l-1)$ gadgets $\prod_{k}$
\onergadget{l-1}{g_k}{I_k'}{J_k'}, executable in $c$
cycles, then \onergadget{l}{G}{I}{J} can be
executed in the same $c$ cycles by being decomposed as $\prod_{k}$
\onergadget{l-1}{g_k}{I\otimes I_k'}{J\otimes
J_k'}. An example is illustrated in \cref{fig:mapping}~(b).

\item \tworgadget{l}{G}{I}{J}: If $G$ is composed of $n$
level-$(l-1)$ gadgets $\prod_{k}$
\tworgadget{l-1}{g_k}{I_k'}{J_k'}, executable in $c$
cycles, then \tworgadget{l}{G}{I}{J} can be
executed in the same $c$ cycles by being decomposed as $\prod_{k}$
\tworgadget{l-1}{g_k}{I\otimes I_k'}{J\otimes
J_k'}.
\end{itemize}

To optimize scheduling within each concatenation level, we utilize a
greedy batching algorithm (illustrated in
\cref{alg:compile_gadgets}), ensuring high parallelism while
maintaining legality constraints.

\subsection{Key Properties and Guarantees of \vair{}}

By iteratively applying the mapping and lowering steps enabled by
\vair{}, we achieve two critical properties for the compiled schedules:

\begin{itemize}
\item \textbf{Level-wise legalization ensures global legalization:}
Any level-$l$ gadget expressed in \vair{} guarantees native
executability on neutral atom arrays once compiled down to physical
instructions.
\item \textbf{Level-wise optimization ensures global optimization:}
Optimizing schedules at level-$l$ directly improves their compiled
physical-level implementations, ensuring monotonic performance gains.
\end{itemize}

These properties empower \name{} to efficiently compile concatenated
codes into physically realizable instructions, significantly
outperforming conventional compilation approaches.

\section{Evaluation}

We present a comprehensive evaluation of \name{}. First, we compare its
performance against state-of-the-art baselines on key fault-tolerant gadgets
and logical-level subroutines. Subsequently, we conduct an ablation study to
quantify the individual contributions of the two main innovations in \name{}:
the \aha{} \cx{} gate scheme and the compilation framework based on \vair{}.

\subsection{Evaluation Methodology}

\noindent\textbf{Baselines.} Our evaluation examines three distinct
compilers: Atomique~\cite{wang2024atomique},
Enola~\cite{tan2025compilation}, and \name{}. For \cx{} gate schemes, we
evaluate both distillation-based approaches and our \aha{} scheme.
All compilation experiments
were conducted on a machine equipped with an AMD EPYC 7763 CPU and
256GB RAM with a time limit of 24 hours. For both Atomique and
Enola, since the default configuration can't complete most of our
benchmarks within the time limit, we use their most scalable configurations.

\noindent\textbf{Benchmarks.} Our evaluation focuses on three
representative fault-tolerant gadgets: (1) \emph{state preparation},
the fundamental building block for initializing logical qubits; (2)
\emph{logical $\cx$ gates}, essential for universal quantum
computation; and (3) \emph{16-qubit GHZ state preparation}, which probes
multi-qubit entanglement and synchronization capabilities. We
evaluate these benchmarks
across four different code configurations: $D_{4,4}$, $D_{4,4,4}$,
$D_{4,4,4,4}$, and $D_{4,4,6,6}$\footnote{We choose
  $D_{4,4,6,6}$ instead of $D_{6,6,4,4}$ because $D_4$ codes have
  better threshold than $D_6$ codes, and concatenating codes with
  higher threshold in the lower levels is more
advantageous~\cite{yoshida2024concatenate,yamasaki2024time}.}.
For comparing \mhcc{} with alternative
quantum QEC codes, we employ the memory experiment, which measures
the logical error rate after one round of error correction as a function of the
physical error rate~\cite{acharya2024quantum}.

\begin{table}[t]
    \centering
    \begin{tabular}{ccccc}
            \toprule
$e_{\text{1Q}}$&$e_{\text{2Q}}$&$e_{\text{move}}$&$e_{\text{reset}}$&$e_{\text{meas}}$\\
    \midrule
    0.03\%&0.5\%&0.1\%&0.25\%&0.25\%\\
    \bottomrule

    \end{tabular}
\caption{Error rates used in our neutral atom array simulation model, based on recent experimental results~\cite{bluvstein2022quantum}.}
\label{tab:parameters}
\vspace*{-0.5cm}
\end{table}

\noindent\textbf{Error Model.} For
the memory experiment, we consider five primary error sources in neutral
atom arrays: two-qubit gate errors ($e_{\text{2Q}}$), single-qubit gate errors
($e_{\text{1Q}}$), atom movement errors ($e_{\text{move}}$), qubit reset errors
($e_{\text{reset}}$), and measurement errors ($e_{\text{meas}}$). Given the
long coherence times of neutral atoms, decoherence errors are negligible and
thus omitted from our model. We use the concrete error rates reported in~\cite{bluvstein2022quantum}, summarized in~\cref{tab:parameters}. For
experiments requiring a range of physical error rates, we treat the two-qubit
gate error rate as the reference point and scale all other error sources
proportionally according to \cref{tab:parameters}.

\noindent\textbf{Simulation Tools.} Surface code simulations utilize
the open-source Stim framework~\cite{gidney2021stim} in conjunction
with pyMatching~\cite{higgott2023sparse} for logical error rate
estimation. For the many-hypercube code, which leverages a concatenated
construction, we developed a specialized simulator that exploits the
underlying hierarchical structure, enabling accelerated sampling.

\noindent\textbf{Decoding.} We implement the level-by-level decoding
approach proposed in~\cite{goto2024many}. This decoding algorithm iteratively
identifies the minimum-distance codeword at lower levels, then
propagates these results to decode codewords at higher levels and has
been demonstrated to outperform both
hard-decision and Bayesian-based decoding methods for concatenated
codes.

\begin{table*}[t]
  \centering
  \small
  \resizebox{\textwidth}{!}{
    \begin{tabular}{ccccccccccccc}
      \toprule
      \textbf{Compiler} &\multicolumn{4}{c}{\textbf{Atomique}} &
      \multicolumn{4}{c}{\textbf{Enola}} &
      \multicolumn{4}{c}{\textbf{\name{}}}  \\
      \cmidrule(lr){2-5}
      \cmidrule(lr){6-9}
      \cmidrule(lr){10-13}
      \textbf{CNOT Scheme}  & \multicolumn{2}{c}{Distillation}  &
      \multicolumn{2}{c}{\aha{}}  &
      \multicolumn{2}{c}{Distillation} & \multicolumn{2}{c}{\aha{}} & \multicolumn{2}{c}{Distillation} &
      \multicolumn{2}{c}{\aha{}} \\
      \cmidrule(lr){2-3}
      \cmidrule(lr){4-5}
      \cmidrule(lr){6-7}
      \cmidrule(lr){8-9}
      \cmidrule(lr){10-11}
      \cmidrule(lr){12-13}
      \textbf{Metric}   & \makecell{Comp.
      \\Time}&\makecell{S.T.\\Prod.}&\makecell{Comp.\\Time}&\makecell{S.T.\\Prod.}&\makecell{Comp.\\Time}&\makecell{S.T.\\Prod.}&\makecell{Comp.\\Time}&\makecell{S.T.\\Prod.}&\makecell{Comp.\\Time}&\makecell{S.T.\\Prod.}&\makecell{Comp.\\Time}&\makecell{S.T.\\Prod.}\\
      \midrule
      State
      Prep($D_{4,4}$)&3.9&0.006&3.9&0.006&0.06&0.004&0.06&0.004&0.0001&0.0007&0.0001&0.0007\\
      State
      Prep($D_{4,4,4}$)&18.7&0.3&18.7&0.3&0.4&0.1&0.4&0.1&0.0005&0.007&0.0005&0.007\\
      State
      Prep($D_{4,4,4,4}$)&4840.7&40.4&4840.7&40.4&22.1&34.7&22.1&34.7&0.001&0.1&0.001&0.1\\
      State
      Prep($D_{4,4,6,6}$)&7695.9&94.5&7695.9&94.5&50.9&87.6&50.9&87.6&0.001&0.2&0.001&0.2\\
      CNOT($D_{4,4}$)&43.3&2.2&62.1&2.7&1.6&1.4&1.5&1.3&0.002&0.05&0.0007&0.03\\
      CNOT($D_{4,4,4}$)&78185.2&288.5&10888.4&146.3&147.0&239.0&68.1&118.6&0.008&1.1&0.005&0.3\\
      CNOT($D_{4,4,4,4}$)&T.O.&T.O.&T.O.&T.O.&12508.0&17820.8&3999.6&6700.1&0.03&20.0&0.04&4.3\\
      CNOT($D_{4,4,6,6}$)&T.O.&T.O.&T.O.&T.O.&73371.1&78612.1&9908.7&15002.6&0.04&54.6&0.06&6.8\\
      GHZ($D_{4,4,4,4}$)&T.O.&T.O.&T.O.&T.O.&50561.1&72188.2&529.3&905.0&0.1&83.2&0.01&3.1\\
      GHZ($D_{4,4,6,6}$)&T.O.&T.O.&T.O.&T.O.&T.O.&T.O.&3403.1&11156.5&0.2&227.6&0.02&9.2\\

      \bottomrule
    \end{tabular}
  }
  \caption{Performance comparison of different compilers and different $\cx$ gate implementation schemes. Results show compilation time (Comp. Time, in seconds) and space-time overhead (S.T. Prod., in $10^6$ qubit-cycles) across different codes and benchmark circuits. ``T.O.'' indicates timeout ($>24$ hours). Note: the overhead of state preparation does not depend on the $\cx{}$ implementation scheme.
  }
  \label{tab:main}
  \vspace*{-0.4cm}
\end{table*}

\subsection{\name{}'s Overall Performance}

\cref{tab:main} summarizes the performance of \name{}
against the baselines, demonstrating that \name{} significantly outperforms
conventional compilers across all metrics. For state preparation on $D_{4,4,6,6}$, compilation
times decrease from several hours with Atomique or $50.9$
seconds with Enola to merely $0.001$ seconds with \name{}. While
Atomique fails to compile level-4 \cx{} and
GHZ circuits within the 24-hour time limit, \name{} consistently completes these
compilations within $0.2$ seconds, and reduces the spacetime product
by up to $2\times 10^3$ ($\cx$ gate on $D_{4,4,6,6}$) compared to Enola.
Moreover, our \aha{} \cx{} gate scheme reduces
the spacetime product by up to $8\times$ for a single \cx{} gate and up
to $20\times$ for GHZ state preparation.

These substantial performance gains stem from two core innovations in \name{}:
efficient addressable gates enabled by the \aha{} gate scheme and
parallelism-preserving scheduling enabled by the \vair{}. We
analyze these two
components in greater detail in the following sections.

\subsection{Impact of \aha{} \cx{} Gate Scheme and Scheduling with \vair{}}

Since Enola consistently outperforms Atomique, we focus our
comparative analysis on \name{} with Enola.

\begin{figure}[t]
  \centering
  \includegraphics[width=\columnwidth]{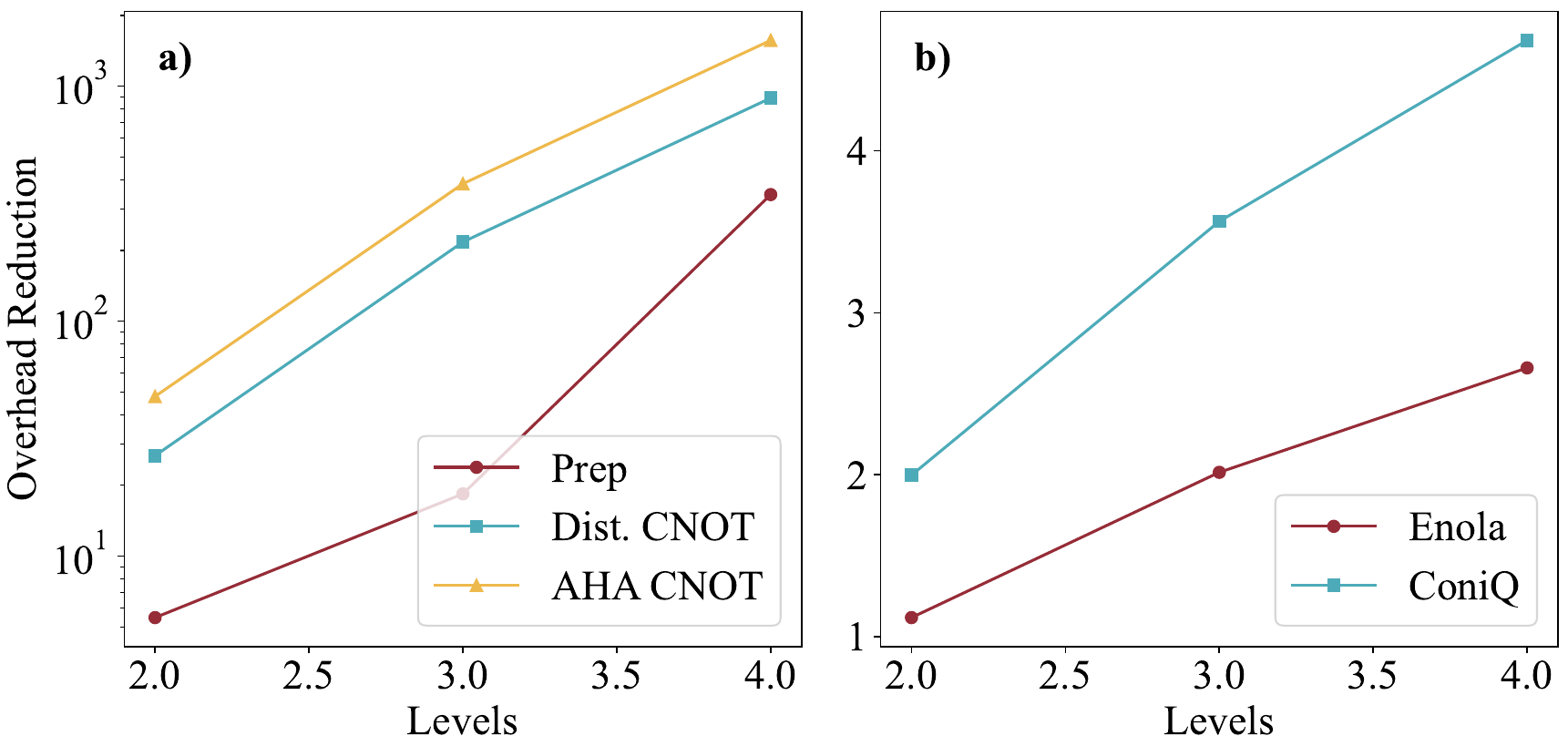}
  \caption{Overhead reduction factors (higher is better) on concatenated $D_4$ codes: (a) Spacetime overhead reduction achieved by \vair{} compared to Enola across concatenation levels. (b) Efficiency comparison between \aha{} and distillation-based $\cx{}$ gate implementations.}
  \label{fig:ablation}
  \vspace*{-0.5cm}
\end{figure}

\noindent\textbf{Level-wise Scheduling.}
\name{}'s \vair{} framework enables level-wise optimization and
legalization that effectively mitigates cascading latency
amplification. As shown in \cref{fig:ablation}~(a), space-time reduction
grows significantly with increasing concatenation levels, reaching
$10^3$ at level-4. The nearly straight line in logarithmic scale
confirms that \name{} successfully avoids the multiplicative overhead
that would otherwise compound across levels.

\noindent\textbf{\aha{} \cx{} Gate Scheme.} Our \aha{} scheme maps
logical \cx{} operations directly to controlled atom movements
compatible with row-column addressability constraints. This reduces
overhead compared to distillation-based approaches, yielding up to
$8\times$ reduction in spacetime product for \cx{} operations and
$20\times$ for GHZ state preparation, with benefits increasing at
higher concatenation levels. \cref{fig:ablation}~(b) shows the \aha{} scheme performs better with \name{}
than Enola because \name{} can better exploit the higher parallelism
of transversal gates.

Together, the \aha{} scheme and parallelism-preserving scheduling
deliver substantial resource savings, demonstrating the importance of
hardware-aware optimizations for scalable quantum error correction on
neutral atom arrays.

\subsection{Discussion: \MHCC{} vs. Other Codes}

\begin{figure}[t]
    \centering
    \includegraphics[width=\columnwidth]{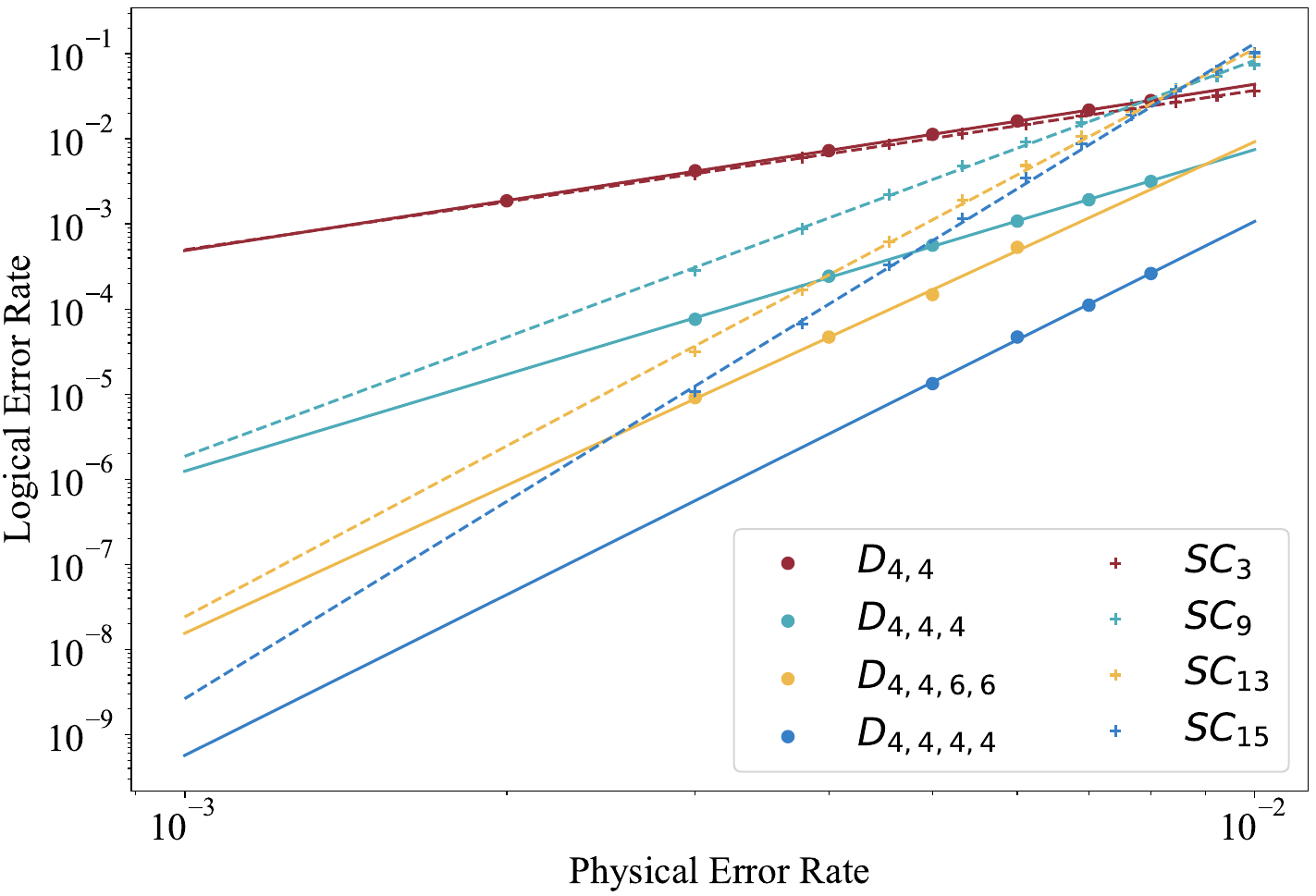}
    \caption{Logical error rate as a function of physical error rate. $SC_d$ denotes surface codes with distance $d$.}
    \label{fig:logical_error_rate}
    \vspace*{-0.5cm}
\end{figure}

We further assess the performance of \mhcc{} by comparing their logical error
rate and space overhead against other leading candidate codes for neutral atom
arrays.

\noindent\textbf{Comparison with Surface Codes.}
\cref{fig:logical_error_rate} compares logical error rates of \mhcc{}
and surface code during memory experiments. \Mhcc{} use Steane-style error
correction while the surface codes use Shor-style error correction with $d$
rounds of stabilizer measurements per logical time
step~\cite{huang2024comparing}.

We extrapolate the logical-to-physical error rate relationship as
$p_{\text{l}}=\beta(\frac{p_{\text{ph}}}{p_{\text{th}}})^\alpha$,
where $\alpha$ is the code distance scaling factor, $\beta$ is the
error coefficient, and $p_{\text{th}}$ is the threshold error rate.

At a physical error rate of $0.1\%$, $D_{4,4,4,4}$ achieves better
logical error rates than $SC_{15}$ while using only 16 physical
qubits per logical qubit compared to 225 for the surface code—a
$14\times$ reduction. The $D_{4,4,6,6}$ code achieves an even greater
$19\times$ reduction in physical qubit requirements.

\noindent\textbf{Comparison with HGP Codes.} \cref{fig:comparison_hgp}
compares our $D_{4,4,4,4}$ code with the hypergraph product (HGP)
codes~\cite{xu2024constant}. The HGP codes treat the entire qubit
array as a single logical block
with error rates improving with system size, while $D_{4,4,4,4}$
maintains constant logical error rates using fixed-size blocks of
$4^4$ physical qubits. Though HGP has advantages in scenarios with
lower physical
error rates and high qubit counts, our analysis shows that it
requires over $10^5$ physical
qubits at $0.1\%$ error rate or $5 \times 10^4$ qubits at $0.01\%$ to
outperform $D_{4,4,4,4}$, far exceeding what will be available in the
near future.

\noindent\textbf{Summary.} The \mhcc{} achieves competitive logical
error rates with substantially lower space overhead compared to
surface codes while avoiding the prohibitive qubit requirements of
HGP codes, making it well-suited for near-term neutral atom quantum
devices with limited qubit availability and quality. However, it is
important to note that the relative advantage changes when
implementing logical gates. A comprehensive evaluation of
space-time costs in the context of complete quantum algorithms
remains an important direction for future research, particularly
considering the difference between Shor- and Steane-style
error correction, and new decoding algorithms that achieve effective
single-shot error correction in surface codes should also be
considered~\cite{zhou2024algorithmic}.

\begin{figure}[t]
  \centering
  \includegraphics[width=\columnwidth]{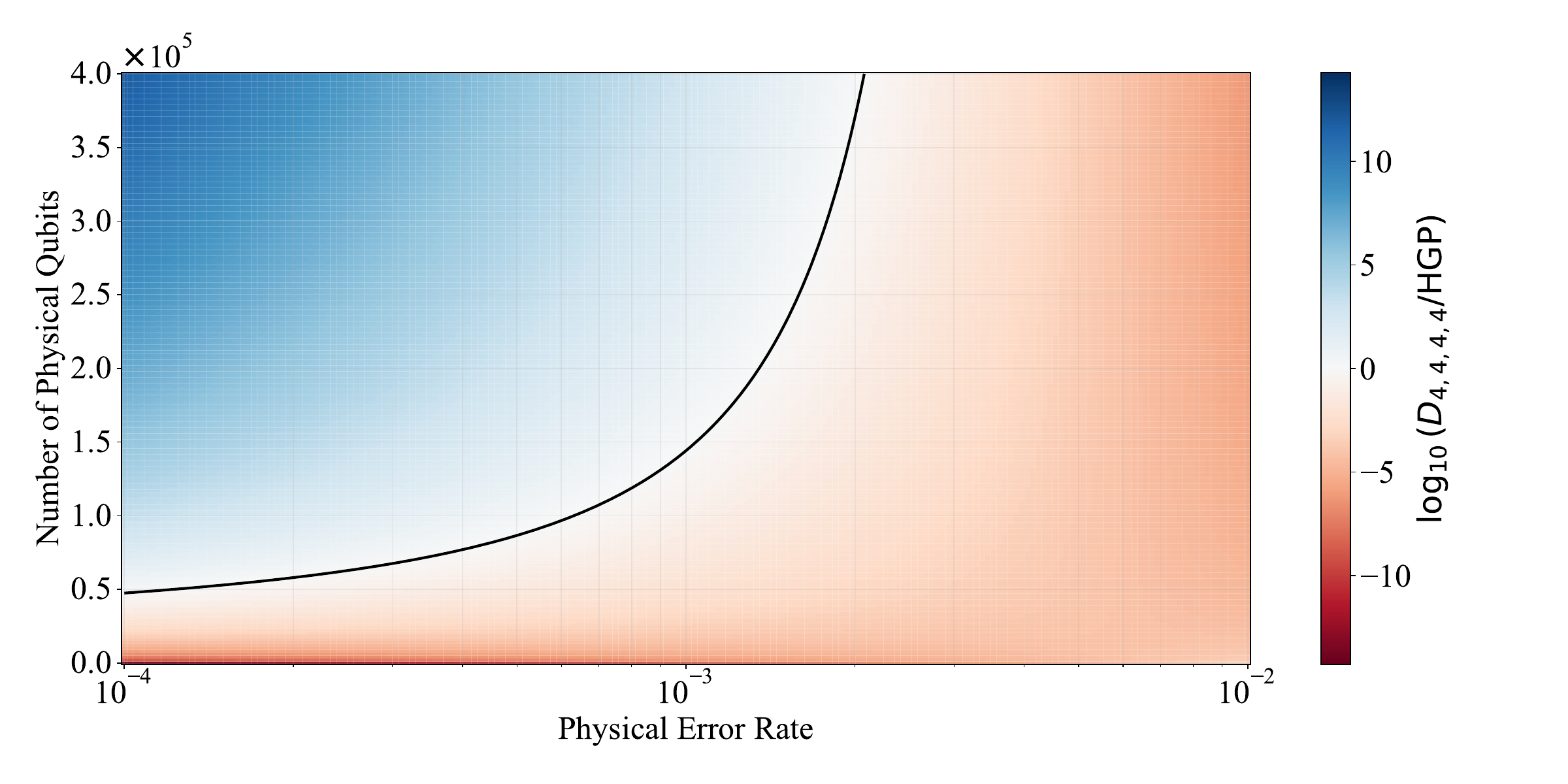}
  \caption{Comparative performance of $D_{4,4,4,4}$ and HGP codes
    under varying physical error rates and qubit counts. Color
    indicates the ratio of logical error rates ($D_{4,4,4,4}$ to HGP).
    Blue region indicates HGP outperforms $D_{4,4,4,4}$ and the black
    line indicates the boundary where the two codes have the same
  logical error rate.}
  \label{fig:comparison_hgp}
  \vspace*{-0.5cm}
\end{figure}

\section{Related Work}
\noindent\textbf{Compilers for Neutral Atom Arrays.} Several
compilers for neutral atom arrays have been
proposed~\cite{baker2021exploiting, patel2022geyser,
tan2022,wang2024atomique,wang2023q,tan2025compilation}. These compilers could
be seamlessly integrated into the \vair{} framework to enhance the simple
greedy algorithm currently employed in \name{} for optimizing register mapping
and scheduling. Additionally, the state preparation scheme of \mhcc{} relies on
post-selection, which bears significant similarities to atom loss management in
neutral atom arrays. Techniques developed for handling atom
loss~\cite{baker2021exploiting,patel2022geyser} may provide valuable insights
for efficiently managing post-selection in our context. Further, the
parallel control of multiple physical qubits to implement parallel
logical operations demonstrated in Ref.~\cite{bluvstein2024logical}
can also be viewed as a special case of the \vair{} model.

\noindent\textbf{Compilers for Fault-Tolerant Quantum Computing.} Prior
research on fault-tolerant quantum computing compilers has predominantly
focused on surface codes, addressing diverse aspects including
resource estimation~\cite{beverland2022assessing}, compilation for trapped
ion~\cite{leblond2023tiscc} and
superconducting~\cite{watkins2024high,molavi2023compilation} architectures.

\noindent\textbf{Concatenated QEC Codes.} Concatenated codes have a rich
history in fault-tolerant quantum computing, and our compilation
framework can be readily applied to various concatenated code
constructions. By concatenating the 7-qubit Steane code with the 15-qubit Reed-Muller code,
Ref.~\cite{chamberland2016thresholds} demonstrates a 105-qubit code capable of
performing $\cx{}$ gates efficiently and implementing $\hgate$ and
$\tgate$ gates
with relative ease. Through concatenation of high-threshold codes with
high-rate codes, Ref.~\cite{yoshida2024concatenate,yamasaki2024time}
achieves a $2.4\%$
threshold ($7$ times higher than the surface code) while requiring $10$ times
less space overhead than conventional surface codes.
Ref.~\cite{pattison2023hierarchical,gidney2023yoked} introduced hierarchical codes and yoked surface code respectively, which
concatenate surface code with other codes,
enabling high code rates with 2D
topological constraints.

\section{Conclusion and Outlook}

In this paper, we have presented \name{}, a specialized compiler for
efficiently implementing concatenated QEC codes on neutral atom
arrays. Through the
\vair{} model and the \aha{} gate scheme, we effectively manage the
complex constraints
imposed by neutral atom array architectures 
and achieve orders of magnitude improvement in
both spacetime overhead and compilation time compared to state-of-the-art
compilers. These results establish the \mhcc{} as a promising candidate for
fault-tolerant quantum computation in the near future.

While our current evaluation focuses primarily on Clifford circuits, our
compiler framework can be straightforwardly extended to support arbitrary
quantum circuits through consuming magic states.

\section*{Acknowledgments}
This research was supported by the following NSF grants CCF-1901381,
CCF-2115104, CCF-2119352, CCF-2107241.  We are grateful to Chameleon
Cloud for providing the compute cycles needed for the experiments.

\bibliographystyle{IEEEtranS}
\bibliography{ref}

\end{document}